\documentclass[aps,prl,twocolumn,superscriptaddress,showpacs,amsmath,amssymb,letterpaper]{revtex4}

\ifx\pdfoutput\@undefined\usepackage[usenames,dvips]{color}
\else\usepackage[usenames,dvipsnames]{color}
\usepackage{graphicx}
\usepackage{bm}
\usepackage{amssymb}
\usepackage{hyperref}
\usepackage{color}
\begin{document}

\title{Magnetoelectric Effects in Local Light-Matter Interactions}


\author{Konstantin Y. Bliokh}
\affiliation{iTHES Research Group, RIKEN, Wako-shi, Saitama 351-0198, Japan}
\affiliation{CEMS, RIKEN, Wako-shi, Saitama 351-0198, Japan}

\author{Yuri S. Kivshar}
\affiliation{Nonlinear Physics Center, Research School of Physics and Engineering, Australian National University,
Canberra ACT 0200, Australia}

\author{Franco Nori}
\affiliation{CEMS, RIKEN, Wako-shi, Saitama 351-0198, Japan}
\affiliation{Physics Department, University of Michigan, Ann Arbor, Michigan 48109-1040, USA}

\begin{abstract}
We study the generic interaction of a monochromatic free-space electromagnetic field with a bi-isotropic nanoparticle. Such interaction is described by dipole-coupling terms associated with the breaking of dual, P- and T-symmetries, including chirality and nonreciprocal magnetoelectric effect. We calculate absorption rates, radiation forces, and radiation torques for such nanoparticles and introduce novel characteristics of the field quantifying the transfer of energy, momentum, and angular-momentum in these interactions. In particular, we put forward a  concept of `magnetoelectric energy density', quantifying the local PT-symmetry of the field. Akin to the `super-chiral' light suggested recently for local probing of molecular chirality, here we suggest to employ complex fields for a sensitive probing of the nonreciprocal magnetoelectric effect in nanoparticles or molecules.
\end{abstract}


\maketitle

\textit{Introduction.---}
Light-matter interactions play a crucial role in various measurement and manipulation processes.
Modern trends imply miniaturization of matter and the use of complex  (i.e., structured, non-plane-wave) fields. Employing local interactions of {\it complex} light with {\it small} particles, investigations of the past decade resulted in: breakthrough in optical manipulations \cite{2}; measurements of angular momentum \cite{3} and nontrivial momentum \cite{Berry,4} of light; reconstruction of particle and complex-field properties at the nanoscale \cite{5}; and sensitive probing \cite{6,8,Chirality,Cho} and sorting \cite{26} of molecular chirality .

One of the important aspects of light-matter interactions is the relations and coupling between {\it electric} and {\it magnetic} properties of both light and matter. In particular, the interaction of light with chiral particles involves the cross-coupling between electric and magnetic characteristics \cite{6,Cho,7}, and chirality of matter turns out to be coupled to the {\it helicity} of light \cite{8}. Remarkably, although chirality is characterized by broken {\it parity} (P-) {\it symmetry} \cite{7}, the conservation of the helicity is intimately related to the less known {\it dual} (D-) {\it symmetry} between electric and magnetic properties of the system \cite{9,10}. The dual symmetry was recently recognized as an important characteristic in light-matter interactions \cite{10,11}.

The cross-coupling between electric and magnetic properties of light and matter can be of different origins \cite{12}. In addition to chirality, there is an intriguing {\it magnetoelectric} interaction currently attracting enormous attention in solid-state physics \cite{13,14}. Such magnetoelectric effect was discovered in 1960 in anisotropic ${\rm Cr_2O_3}$ \cite{15}, but magnetoelectric materials were still considered to be rare until recently. Importantly, the magnetoelectric effect implies {\it nonreciprocity}: the magnetoelectric matter has broken P- and {\it time-reversal} (T-) {\it symmetries} \cite{12,14} but preserved combined {\it PT-symmetry}, which is now a subject of active research in quantum mechanics and optics \cite{16}.

Nonreciprocal magnetoelectric effects appear in contrasting and sometimes controversial contexts. At the molecular level, it was first discussed by Curie and Debye \cite{Curie}, but has never been detected. Later, Barron \cite{Barron} named it ``{\it false chirality}'', provided the symmetry analysis, and revealed an analogy with the {\it violation of CP-symmetry} in particle physics. For continuous media, in 1948, Tellegen \cite{17} described properties of a hypothetic bi-isotropic magnetoelectric medium, ``{\it Tellegen medium}'' \cite{12}. The possible existence of such medium was actively debated in the 1990s \cite{18}, and it was eventually realized recently for static fields \cite{Tellegen medium}. It was also shown that the magnetoelectric effect is closely related to {\it axion} electrodynamics \cite{19}, and can appear in {\it topological insulators} \cite{20}.

While chirality of matter is naturally coupled to the helicity of light \cite{6,8}, what is the field characteristic that can probe the nonreciprocal magnetoelectric property (``false chirality'') of matter? In this Letter we address this question by considering the dipole interaction of light with a generic bi-isotropic small particle. We describe contributions related to the broken P-, T-, and D- symmetries and define the local energy, momentum, and angular-momentum characteristics of the field, which quantify these specific interactions. In particular, we introduce a PT-symmetric {\it `magnetoelectric energy density'}, which is coupled to the magnetoelectric polarizability of matter. Importantly, this characteristic vanishes for plane waves, but it can be non-zero for {\it complex} fields. Akin to the `super-chiral light' recently suggested for sensitive local probing of molecular chirality \cite{6,Chirality,Cho,8}, we put forward a {\it `super-magnetoelectric'} field for sensitive probing of the magnetoelectric properties of nanoparticles. This can lead to the discovery of the nonreciprocal magnetoelectric effect at the molecular level, as was first discussed about a century ago \cite{14,Curie}.

\textit{Characteristics of monochromatic Maxwell fields.---}
We consider monochromatic Maxwell's electric and magnetic fields, $\bm{\mathcal E}( {\bf{r},t} ) = {\Re}\! \left[ {{\bf{E}}( {\bf{r}} ){e^{ - i\omega t}}} \right]$ and $\bm{\mathcal H}( {\bf{r},t} ) = {\Re}\! \left[ {{\bf{H}}( {\bf{r}} ){e^{ - i\omega t}}} \right]$, in free space, and use electrodynamics units $c = {\varepsilon _0} = {\mu _0} = 1$. Dynamical properties of light can be conveniently described within a quantum-like formalism by introducing a complex two-level `state vector' of electric and magnetic fields \cite{Berry,10}:
\begin{equation}
\label{Eq01}
\vec{\bm \psi} ( {\bf{r}} ) = \frac{1}{{2\sqrt \omega  }}\left( {\begin{array}{*{20}{c}}
{{\bf{E}}( {\bf{r}} )}\\
{{\bf{H}}( {\bf{r}})}
\end{array}} \right)~.
\end{equation}
The state vector (1) is formally a vector in $\mathbb{C}^3 \otimes \mathbb{C}^2 \otimes \mathbb{L}^2$ space, where vectors in usual $\mathbb{C}^3$ space are denoted by bold letters, vectors in the electric-magnetic (dual) $\mathbb{C}^2$ space are denoted by arrows, and vectors in Hilbert $\mathbb{L}^2$ space are functions of $\bf{r}$ \cite{remark}.

Stationary free-space Maxwell equations play the role of constraints for the `state vector' (1), which has too many degrees of freedom in the $\mathbb{C}^3 \otimes \mathbb{C}^2 \otimes \mathbb{L}^2$ space. They include the transversality equations and stationary Weyl-like equation for spin-1 photons \cite{23}:
\begin{equation}
\label{Eq02}
{\bf{\hat p}} \cdot \vec {\bm \psi} ( {\bf{r}} ) = 0~,~~~
\left( {\frac{{{\bf{\hat S}} \cdot {\bf{\hat p}}}}{\omega }} \right)\vec {\bm \psi} ( {\bf{r}} ) =  - {\hat \sigma _2}\,\vec {\bm \psi} ( {\bf{r}} )~.
\end{equation}
Here ${\bf{\hat p}} =  - i{\bm \nabla}$ is the momentum operator that acts in $\mathbb{L}^2$, ${\bf{\hat S}}$ is the vector of spin-1 rotational matrices which operate in $\mathbb{C}^3$ as ${{\bf{X}}^*} \!\cdot\! ( {{\bf{\hat S}}} ){\bf{X}} \!=\! {\Im}\!\left( {{{\bf{X}}^*}\! \times \!{\bf{X}}} \right)$, and the Pauli matrices $\hat\sigma_i$, $i=1,2,3$, act in the $\mathbb{C}^2$ `electric-magnetic' space with the unit matrix ${\hat \sigma _0} = {\rm diag} \left( {1,1} \right)$ being implied if there is no $\hat\sigma_i$.
Throughout the paper we adopt the notation \cite{Berry}: ${{\bf{X}}} \!\cdot\! ( {{\bf{\hat Z}}} ){\bf{Y}} \!=\! \sum_{i} X_i {{\bf{\hat Z}}} {Y_i}$.
The second Eq.~(2) shows that the operator $\hat K =  - {\hat \sigma _2}$ acting on the Maxwell fields is equivalent to the quantum helicity operator ${\bf{\hat S}} \cdot {\bf{\hat p}}/p$ ($p = k = \omega$). The operator $\hat K$ is a generator of the {\it dual transformation} $\exp ( { - i \theta \hat{K}} )$ rotating the electric and magnetic properties \cite{9,10,11}, and its eigenmodes $\hat K \vec {\bm \psi}  = \chi \vec {\bm \psi}$ with $\chi=\pm 1$ are the fields with well-defined {\it helicity} $\chi$: ${\bf{H}} =  - i\chi {\bf{E}}$. This illuminates the connection between the helicity and duality in Maxwell fields.

The main local dynamical characteristics of the electromagnetic field are: energy, momentum, angular-momentum (spin), and helicity densities. These quantities, averaged over time, can be written using Eq.~(1) as local expectation values of the corresponding operators \cite{Berry,10}:
\begin{eqnarray}
\label{Eq03-06}
W\!= {\vec {\bm \psi} ^\dag }\!\cdot\!\left( \omega  \right) \vec {\bm \psi} = \frac{1}{4}\left( {{{\left| {\bf{E}} \right|}^2} + {{\left| {\bf{H}} \right|}^2}} \right),\\
{\bf{P}}\!=\!{\Re} \!\left[ {{{\vec {\bm \psi} }^\dag }\!\cdot\!\left( {{\bf{\hat p}}} \right) \vec {\bm \psi} } \right]\!=\!\frac{1}{{4\omega }}{\Im}\! \left[ {{{\bf{E}}^{*}}\!\cdot\!\left( \bm{\nabla}  \right){\bf{E}} + {{\bf{H}}^{\ast}}\!\cdot\!\left( \bm{\nabla}  \right){\bf{H}}} \right],\\
{\bf{S}} = {\vec {\bm \psi} ^\dag }\! \cdot\! \left( {{\bf{\hat S}}} \right)\!\vec {\bm \psi}  = \frac{1}{{4\omega }}{\Im} \!\left( {{{\bf{E}}^*}\! \times {\bf{E}} + {{\bf{H}}^*}\! \times {\bf{H}}} \right),\\
K \!= {\vec {\bm \psi} ^\dag } \!\cdot\! \left( {\hat K} \right)\!\vec {\bm \psi} =  - \frac{1}{{2\omega }}{\Im}\! \left( {{{\bf{E}}^*} \!\cdot {\bf{H}}} \right).
\end{eqnarray}
Here $\bf P$ is the {\it canonical} or orbital momentum density that describes optical pressure, energy transport, and appears in experiments \cite{Berry,4,10}. It also determines the density of orbital angular momentum ${\bf{L}} = {\bf{r}} \times {\bf{P}}$ \cite{3,4}, which is not an independent quantity, in contrast to the spin density $\bf S$. The widely used (but, unlike $\bf{P}$, not directly measurable \cite{4}) Poynting vector ${\bm \Pi}$ is obtained by adding the virtual `spin momentum' to $\bf P$ (for details, see \cite{Berry,4,10,24}):
\begin{equation}
\label{Eq07}
{\bm \Pi} = {\vec {\bm \psi} ^\dag }\! \cdot \!\left( {\omega \hat K {\bf{\hat S}}} \right) \!\vec {\bm \psi} = {\bf{P}} + \frac{1}{2}{\bm \nabla}\! \times {\bf{S}} = \frac{1}{2}{\Re} \!\left( {{{\bf{E}}^*}\! \times {\bf{H}}} \right),
\end{equation}
where the Maxwell equations (2) were used. Equation (7) also supports the fact that the Poynting vector is not a fundamental physical characteristic because its underlying operator $\omega \hat K {\bf{\hat S}}$ mixes different subspaces. In contrast, the fundamental quantities (3)--(6) are underpinned by operators in the corresponding spaces: momentum in $\mathbb{L}^2$, spin in $\mathbb{C}^3$, and helicity/duality in $\mathbb{C}^2$. Therefore, quantities (3)--(5) represent the sums of separate electric and magnetic parts: $W = {W^{\rm (e)}} + {W^{\rm (m)}}$, ${\bf{P}} = {{\bf{P}}^{\rm (e)}} + {{\bf{P}}^{\rm (m)}}$, ${\bf{S}} = {{\bf{S}}^{\rm (e)}} + {{\bf{S}}^{\rm (m)}}$, while the helicity (6) mixes electric and magnetic properties.

\textit{Interaction with a dipole bi-isotropic particle.---}
We now consider the interaction of a monochromatic Maxwell field (1) with a generic bi-isotropic dipole particle. The particle is characterised by oscillating electric and magnetic dipole moments $\bm{\mathcal{P}}( {{\bf{r}},t} ) = {\Re}\!\left[ {\bm{\pi} ( {\bf{r}} ){e^{ - i\omega t}}} \right]$ and $\bm{\mathcal{M}}( {{\bf{r}},t} ) = {\Re}\!\left[ {\bm{\mu} ( {\bf{r}} ){e^{ - i\omega t}}} \right]$, induced by the field. The dipole moments $\bm{\pi}$ and $\bm{\mu}$ are proportional to the fields $\bf E$ and $\bf H$ with complex scalar coefficients. Hence, the generic form of these relations can be written using the Pauli-matrix expansion of the interaction operator $\hat{\Delta}$ acting in the dual $\mathbb{C}^2$ space:
\begin{eqnarray}
\label{Eq08}
\left( {\begin{array}{*{20}{c}}
\bm{\pi} \\
\bm{\mu}
\end{array}} \right) = \hat \Delta \left( {\begin{array}{*{20}{c}}
{\bf{E}}\\
{\bf{H}}
\end{array}} \right),~\nonumber\\
\hat \Delta  = \left( {{\alpha ^{(0)}}{{\hat \sigma }_0} + \color{red} {\alpha ^{\rm (me)}}{{\hat \sigma }_1} \color{black} + \color{ForestGreen} {\alpha ^{\rm (ch)}}{{\hat \sigma }_2} \color{black} + \color{blue} {\alpha ^{\rm (da)}}{{\hat \sigma }_3}} \color{black} \right),
\end{eqnarray}
where ${\alpha ^{\rm (0,da,ch,me)}}$ are complex polarizabilities describing interactions with different symmetry properties. Namely: $\alpha^{(0)}$ describes excitation of the electric and magnetic dipoles by the corresponding fields with the same efficiency (an {\it `ideal'} dual-symmetric particle without electric-magnetic cross-coupling); $\alpha^{\rm (da)}$ characterizes the difference between the electric and magnetic dipole excitations by the corresponding fields (i.e., the {\it dual asymmetry} between electric and magnetic properties of the particle) \cite{10,11}; $\alpha^{\rm (ch)}$ quantifies {\it chirality} (reciprocal cross-excitation of the electric and magnetic dipoles) [6,7]; and, finally, $\alpha^{\rm (me)}$ describes the {\it magnetoelectric effect} or ``false chirality'' ({\it nonreciprocal} cross-excitation of the electric and magnetic dipoles) \cite{12,13,14,15,Curie,17,18,19,20}.

Light-particle interactions can be characterized by the energy, momentum, and angular-momentum transfer from light to matter. These are quantified by the absorption rate $A$, radiation force $\bf F$, and radiation torque $\bf T$, respectively. Calculations of these time-averaged quantities result in the following equations, conveniently represented in our formalism:
\begin{eqnarray}
\label{Eq09-11}
A =\! - \frac{\omega }{2}{\Im}\! \left[ {{\bm{\pi}^*}\! \cdot {\bf{E}} + {\bm{\mu}^*}\! \cdot {\bf{H}}} \right] =\! \gamma {\Im}\! \left[ {{{\vec {\bm \psi} }^\dag }\! \cdot \!\left( {{{\hat \Delta }^\dag }\omega } \right)\!\vec {\bm \psi} } \right]\!,\\
{\bf F} \!=\! \frac{1}{2}{\Re}\! \left[ {{\bm{\pi}^*}\! \cdot\! \left({\bm{\nabla}}\right)\! {\bf{E}} + {\bm{\mu}^*}\! \cdot \!\left({\bm{\nabla}}\right)\! {\bf{H}}} \right] \!=\! \gamma {\Im}\! \left[ {{{\vec {\bm \psi} }^\dag }\! \cdot \!\left( {{{\hat \Delta }^\dag }{\bf{\hat p}} } \right)\!\vec {\bm \psi} } \right]\!,\\
{\bf T} \!=\! \frac{1}{2}{\Re}\! \left[ {{\bm{\pi}^*}\! \times\!  {\bf{E}} + {\bm{\mu}^*}\! \times \! {\bf{H}}} \right] \!=\! \gamma {\Im}\! \left[ {{{\vec {\bm \psi} }^\dag }\! \cdot \!\left( {{{\hat \Delta }^\dag }{\bf{\hat S}} } \right)\!\vec {\bm \psi} } \right]\!,
\end{eqnarray}
where $\gamma=-2\omega$. Thus, the optical absorption rate, force, and torque are directly related to the canonical energy, momentum, and spin operators, combined with the interaction operator (8) ${\hat \Delta}$.

Let us examine effects from various interaction terms in ${\hat \Delta}$. Substituting Eq.~(8) into Eqs.~(9)--(11), the corresponding absorption rates, forces, and torques can be written in a unified form:
\begin{eqnarray}
\label{Eq12-14}
{A^X} = 2\omega {\Im} \!\left( {{\alpha ^X}} \right) {W^X}~,\\
{{\bf{F}}^{X}} = {\Re} \!\left( {{\alpha ^X}} \right) \bm{\nabla} {W^X} + 2\omega {\Im}\! \left( {{\alpha ^X}} \right) {{\bf{P}}^{X}}~,\\
{{\bf{T}}^{X}} = 2\omega {\Im}\! \left( {{\alpha ^X}} \right) {{\bf{S}}^{X}}~.
\end{eqnarray}
Here $X = (0),{\rm (da),(ch),(me)}$ indicates different interactions, and the optical force (13) is split into the gradient and radiation-pressure parts \cite{2,4,25}. The imaginary parts of the polarizabilities $\alpha^X$ characterize dissipation, so that the terms proportional to $\Im\!\left(\alpha^X\right)$ quantify the local energy/momentum/angular-momentum absorbed by the particle in the corresponding $X$-interaction. For an ideal interaction without any asymmetries, $X=(0)$, Eqs.~(12)--(14) exhibit the fundamental field characteristics (3)--(5): $W^{(0)}=W$, ${\bf P}^{(0)}={\bf P}$, and ${\bf S}^{(0)}={\bf S}$. Thus, the absorption rate, radiation-pressure force, and torque for the ideal particle allow straightforward measurements of the energy, momentum, and spin densities in the field \cite{3,Berry,4}.

For interactions with various asymmetries, Eqs.~(12)--(14) contain modified characteristics $W^X$, ${\bf P}^X$, and ${\bf S}^X$ quantifying the energy, momentum, and angular-momentum exchange in these interactions. First, the {\it dual asymmetry} of the particle is characterized by the $X={\rm (da)}$ interaction, and the corresponding quantities represent differences between electric and magnetic parts of the canonical characteristics:
\begin{eqnarray}
\label{Eq15}
\color{blue}
\boxed{\color{black}{W^{\rm (da)}} = {W^{\rm (e)}} - {W^{\rm (m)}},~{{\bf{P}}^{\rm (da)}} = {{\bf{P}}^{\rm (e)}} - {{\bf{P}}^{\rm (m)}},~{\rm etc.}}
\end{eqnarray}
Typically, matter is strongly dual asymmetric because of the presence of {\it electric} (but not magnetic) charges. For instance, an electric-dipole particle with electric polarizability $\alpha^{\rm (e)}$ and negligible magnetic polarizability is described by ${\alpha ^{\rm (da)}} = {\alpha ^{(0)}} = {\alpha ^{\rm (e)}}/2$ \cite{remark II}.

Second, the {\it chirality} of the particle (P-{\it asymmetry}) is described by the $X={\rm (ch)}$ interaction, and the corresponding `chiral' energy, momentum, and spin densities are given by
\begin{eqnarray}
\label{Eq16-18}
\color{ForestGreen}
\boxed{\color{black}{W^{\rm (ch)}} = \frac{1}{2}{\Im}\! \left( {{{\bf{E}}^*}\! \cdot {\bf{H}}} \right) =  - \omega K}\color{black}~,\\
\color{ForestGreen}
\boxed{\color{black}{{\bf{P}}^{\rm (ch)}} =  - \frac{1}{{4\omega }}{\Re}\! \left[ {{{\bf{E}}^*}\! \cdot \left( \bm{\nabla}  \right){\bf{H}} - {{\bf{H}}^*}\! \cdot \left( \bm{\nabla}  \right){\bf{E}}} \right]} \nonumber\\
=  - \omega {\bf{S}} + \frac{1}{{2\omega }} \bm{\nabla}  \times \bm{\Pi}~,\\
\color{ForestGreen}
\boxed{\color{black}{{\bf{S}}^{\rm (ch)}} =  - \frac{1}{{2\omega}}{\Re}\! \left( {{{\bf{E}}^*}\! \times {\bf{H}}} \right) =  - \frac{\bm{\Pi}}{\omega }}\color{black}~,
\end{eqnarray}
where the Maxwell equations (2) were used in Eq.~(17). Naturally, these characteristics contain mixed electric and magnetic field properties. Equation (16) shows that the absorption rate in the chiral interaction is proportional to the helicity of light $K$ \cite{6,8}, whereas Eqs.~(17) and (18) agree with very recent results of \cite{26} calculating chiral optical forces and suggesting optical sorting of chiral particles. Interestingly, the Poynting vector (7) determines the chiral torque (18), whereas the spin determines one part of the chiral force (17).

Finally, the most intriguing and unexplored case is the {\it magnetoelectric interaction} $X={\rm (me)}$ (P- {\it and} T-{\it asymmetric}). Akin to Eqs.~(16)--(18), we derive that the local magnetoelectric energy, momentum, and angular-momentum exchange are characterized by the following quantities:
\begin{eqnarray}
\label{Eq19-21}
\color{red}
\boxed{\color{black}{W^{\rm (me)}} = \frac{1}{2}{\Re}\! \left( {{{\bf{E}}^*}\! \cdot {\bf{H}}} \right)}~\color{black},\\
\color{red}
\boxed{\color{black}{{\bf{P}}^{\rm (me)}} =-\frac{1}{{4\omega }}\bm{\nabla}\!\times\!{\Im}\! \left( {{{\bf{E}}^*}\! \times {\bf{H}}} \right)} \color{black}~,\\
\color{red}
\boxed{\color{black}{{\bf{S}}^{\rm (me)}} = \frac{1}{{2\omega }}{\Im}\! \left( {{{\bf{E}}^*}\! \times {\bf{H}}} \right)}\color{black}~.
\end{eqnarray}
Remarkably, equation (19) introduces the {\it `magnetoelectric energy density'} $W^{\rm (me)}$, which {\it vanishes in any plane wave} with ${\bf{H}} \propto {\bf{k}} \times {\bf{E}}$. This is because the eigenstates of the magnetoelectric operator $\hat \sigma_1$ are collinear electric and magnetic fields, ${\bf{H}} =  \pm {\bf{E}}$, impossible in plane waves. Nonetheless, such magnetoelectric fields can exist {\it locally} as a result of interference of several plane waves. Thus, {\it the magnetoelectric effect can be locally sensed via absorption rates by specially-designed complex fields with} ${\bf{H}} \parallel {\bf{E}}$ \cite{remark III}.
\begin{widetext}

\begin{table}
\begin{tabular}{ccccc}
\hline
\hline
 & \multicolumn{1}{c}{Fundamental} &
\multicolumn{1}{c}{Dual-asymmetric} &
\multicolumn{1}{c}{Chiral} &
\multicolumn{1}{c}{Magnetoelectric} \\
& \multicolumn{1}{c}{~~~~$W$~~~~~$\bf P$~~~~~$\bf S$~~~~~} &
\multicolumn{1}{c}{~~~$W^{\rm (da)}$~~${\bf P}^{\rm (da)}$~~${\bf S}^{\rm (da)}$~~} & \multicolumn{1}{c}{~~~$W^{\rm (ch)}$~~~${\bf P}^{\rm (ch)}$~~~${\bf S}^{(\rm ch)}$~} & \multicolumn{1}{c}{~~~~$W^{\rm (me)}$~~~${\bf P}^{\rm (me)}$~~~${\bf S}^{\rm (me)}$~} \\ \hline
{P-symmetry}~~ & ~~$+$~~~~~$-$~~~~~$+$~~ & $+$~~~~~~~$-$~~~~~~$+$~~ & ~~$\color{ForestGreen} \bm{-}$~~~~~~~$\color{ForestGreen} \bm{+}$~~~~~~$\color{ForestGreen} \bm{-}$~~ & ~~$\color{red} \bm{-}$~~~~~~~~$\color{red} \bm{+}$~~~~~~~~$\color{red} \bm{-}$~~ \\
{T-symmetry}~~ & ~~$+$~~~~~$-$~~~~~$-$~~ & $+$~~~~~~~$-$~~~~~~$-$~~ & ~~${+}$~~~~~~~${-}$~~~~~~~${-}$~~ & ~$\color{red} \bm{-}$~~~~~~~~$\color{red} \bm{+}$~~~~~~~~$\color{red} \bm{+}$~~\\
{D-symmetry}~~ & ~~$+$~~~~~$+$~~~~~$+$~~ & $\color{blue} \bm{-}$~~~~~~~$\color{blue} \bm{-}$~~~~~~$\color{blue} \bm{-}$~~ & ~~${+}$~~~~~~~${+}$~~~~~~~${+}$~~ & ~$\color{red} \bm{-}$~~~~~~~~$\color{red} \bm{-}$~~~~~~~~$\color{red} \bm{-}$~~\\
\hline\hline
\end{tabular}
\caption
{{Symmetry properties of the fundamental, dual-asymmetric, chiral, and magnetoelectric characteristics of the field.}}
\label{tab:experiments}
\end{table}

\end{widetext}

It is important to examine the P-, T-, and D-symmetry properties of the free-space and interaction field characteristics (3)--(5) and (15)--(21). The corresponding transformations are given by
\begin{eqnarray}
\label{Eq22}
\left\{\color{ForestGreen}\hat{\rm P},\color{red}\hat{\rm T},\color{blue}\hat{\rm D}\right\}\color{black}\!\left( {\begin{array}{*{20}{c}}
{\bf{E}}\\
{\bf{H}}
\end{array}} \right) \!= \!\left\{ {\left( {\begin{array}{*{20}{c}}
{\color{ForestGreen}\!- {\bf{E}}\!}\\
{\color{ForestGreen}\!\bf{H}\!}
\end{array}} \right)\!,\left( {\begin{array}{*{20}{c}}
{{\color{red}\!{\bf{E}}^*\!}}\\
{\color{red}\!- {\bf{H}}^*\!}
\end{array}} \right)\!,\left( {\begin{array}{*{20}{c}}
{\color{blue}\!\bf{H}\!}\\
{\color{blue}\!-{\bf{E}}\!}
\end{array}} \right)} \right\}\!,
\end{eqnarray}
where we consider the discrete dual transformation $\hat{\rm D} = \exp ( { - i\pi \hat K/2} )$ \cite{9,10,11}. Quantities even and odd under these transformations are marked by ``+'' and ``$-$'' in Table I. One can see that the symmetries of the interaction characteristics (15)--(21) are flipped as compared with the fundamental ones, according to the asymmetry of the corresponding interaction. Namely, the dual-asymmetric quantities have flipped D-symmetry, the chiral characteristics have inverted P-symmetry, and the magnetoelectric characteristics have flipped all P-, T-, and D-symmetries, but preserved PT-symmetry.

\textit{Sensitive probing of magnetoelectric particles.---}
Using the above expressions, one can suggest field configurations sensitive to the specific properties of the particle. Recently, Refs.~\cite{6,Chirality,Cho} demonstrated the ultrasensitive local probing of molecular {\it chirality} using field configurations with high ratio of the helicity density to the electric energy density: $\omega \left| K \right|/{W^{\rm (e)}} \gg 1$. This `super-chiral' response is essentially based on the {\it dual asymmetry} of the molecules (small magnetic polarizability) \cite{8,Cho}. Here we suggest a similar method of ultrasensitive local probing of the {\it magnetoelectric} effect.

We consider a typical dual-asymmetric particle with high electric polarizability $\alpha^{\rm (e)}$, small magnetic polarizability: $\Im\!\left({\alpha ^{\rm (m)}}\right) = \varepsilon\, \Im\!\left({\alpha ^{\rm (e)}}\right)$, $| \varepsilon  | \ll 1$, and weak magnetoelectric polarizability: $\left|\Im\!\left( {{\alpha ^{\rm (me)}}} \right)\right| \ll \left|\Im\!\left( {{\alpha ^{\rm (e)}}}\right) \right|$. Using Eqs.~(3), (12), (15), and (19) with ${\alpha ^{(0)}} = \left( {{\alpha ^{\rm (e)}} + {\alpha ^{\rm (m)}}} \right)/2$, ${\alpha ^{\rm (da)}} = \left( {{\alpha ^{\rm (e)}} - {\alpha ^{\rm (m)}}} \right)/2$, we obtain the absorption rate for the electric and magnetic dipole interactions: ${A^{\rm (e)}} + {A^{\rm (m)}} = \left( {\omega /2} \right) \Im\!\left({\alpha ^{\rm (e)}}\right) ( {{{\left| {\bf{E}} \right|}^2} + \varepsilon {{\left| {\bf{H}} \right|}^2}} )$. In turn, the magnetoelectric absorption rate is determined by the magnetoelectric energy (19): ${A^{\rm (me)}} = \omega\, \Im\!\left({\alpha ^{\rm (me)}}\right) {\Re}\!\left( {{{\bf{E}}^*} \!\cdot {\bf{H}}} \right)$. Thus, the relative magnetoelectric response is characterized by the ratio
\begin{equation}
\label{Eq23}
{\tilde A^{\rm (me)}} = \frac{{2{\Re}\! \left( {{{\bf{E}}^*} \!\cdot {\bf{H}}} \right)}}{{{{\left| {\bf{E}} \right|}^2} + \varepsilon {{\left| {\bf{H}} \right|}^2}}}~.
\end{equation}
A similar quantity maximizing the chiral response in \cite{6,Chirality} is ${\tilde A^{\rm (ch)}} =  - 2{\Im}\!\left( {{{\bf{E}}^*}\! \cdot {\bf{H}}} \right)/( {{{\left| {\bf{E}} \right|}^2} + \varepsilon {{\left| {\bf{H}} \right|}^2}} )$ \cite{Cho}, which is limited as $| {{{\tilde A}^{\rm (ch)}}} | < 2$ in plane waves but can achieve anomalously high values $| {{{\tilde A}^{\rm (ch)}}} |\sim 1/\sqrt \varepsilon  \gg 1$ in complex fields near the {\it electric-field nodes}, ${\left| {\bf{E}} \right|^2} \simeq 0$.

The magnetoelectric response factor (23) {\it vanishes} identically in plane waves, but can achieve the same `super-values' $| {{{\tilde A}^{\rm (me)}}} |\sim 1/\sqrt \varepsilon \gg 1$ in complex fields. To illustrate this, we employ almost the same field configuration as was used in \cite{6}. Namely, we consider two counter-propagating plane waves with the {\it same} amplitudes and circular polarizations of the same direction of absolute rotation:
\begin{eqnarray}
\label{Eq24}
{\bf{E}} \propto \left( {{\bar{\bf{x}}} + i\sigma \bar{\bf{y}}} \right){e^{ikz}} + \left( {\bar{\bf{x}} + i\sigma \bar{\bf{y}}} \right){e^{ - ikz}}~,\nonumber\\
{\bf{H}} \propto \left( {{\bar{\bf{y}}} - i\sigma \bar{\bf{x}}} \right){e^{ikz}} - \left( {\bar{\bf{y}} - i\sigma \bar{\bf{x}}} \right){e^{ - ikz}}~.
\end{eqnarray}
Here $\bar{\bf{x}}$ and $\bar{\bf{y}}$ are the unit vectors of the corresponding axes and $\sigma=\pm 1$ determines the circular-polarization direction. The field (24) forms a standing wave with rotating distribution of real electric and magnetic fields ${\bm {\mathcal E}} ( {{\bf{r}},t} )$ and ${\bm {\mathcal H}} ( {{\bf{r}},t} )$ shown in Fig.~1a. One can see that the electric and magnetic fields are {\it collinear} with each other, exactly as required by the magnetoelectric configuration. Substituting field (24) into Eq.~(23), we obtain the magnetoelectric response factor:
\begin{equation}
\label{Eq25}
{\tilde A^{\rm (me)}} = \sigma \frac{{2\sin\!\left( {kz} \right)\cos \!\left( {kz} \right)}}{{{{\cos }^2}\!\left( {kz} \right) + \varepsilon\, {{\sin }^2}\!\left( {kz} \right)}}~.
\end{equation}
Distribution ${\tilde A^{\rm (me)}}( z )$ is plotted in Fig.~1b, where one can see pronounced resonances in the vicinity of the electric nodes $kz = \pi \left( {n + 1/2} \right)$ ($n$ is an integer). The maximal resonant magnetoelectric response is ${| {{{\tilde A}^{\rm (me)}}} |_{res}} = 1/\sqrt \varepsilon$ and the peak width is $k\delta\!{z_{res}} = 2\sqrt {3\varepsilon } $. For realistic materials with $\varepsilon \sim {10^{ - 4}}$--${10^{ - 6}}$, this yields the enhancement ${| {{{\tilde A}^{\rm (me)}}} |_{res}}\sim 30$--$1000$ \cite{Cho}. Importantly, the magnetoelectric absorption (25) is proportional to the polarization ellipticity $\sigma$. Therefore, switching the polarization between the $\sigma=\pm 1$ modes and measuring the difference in the absorption rates one can measure the magnetoelectric polarizability of the particle \cite{29}.

\begin{figure}[t]
\includegraphics[width=8cm, keepaspectratio]{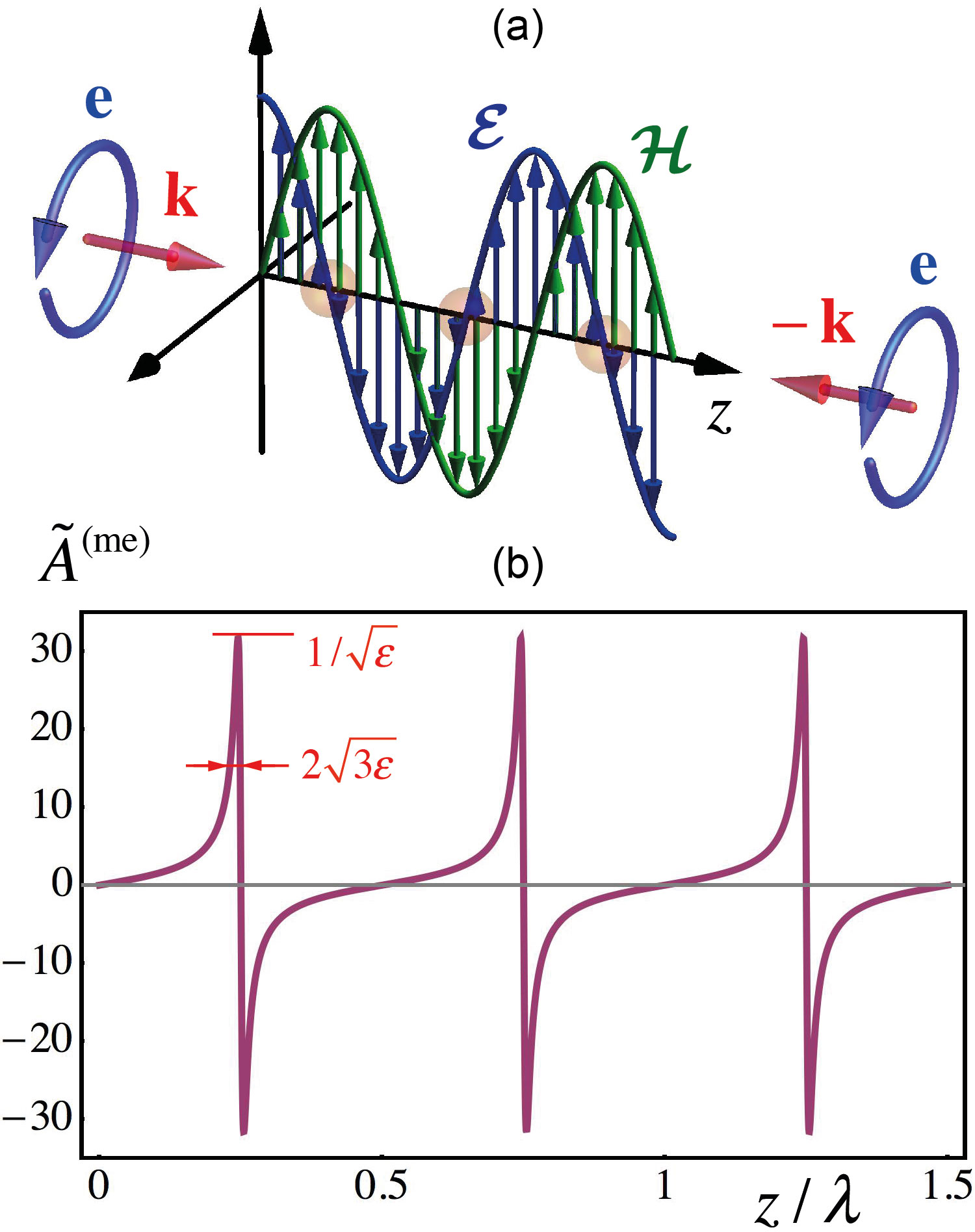}
\caption{(color online) Sensitive probing the magnetoelectric effect in nanoparticles via complex fields. (a) Two counter-propagating circularly-polarized plane waves (24) form a standing wave with rotating collinear electric and magnetic fields. (b) The $z$-dependence of the magnetoelectric response factor (23) and (25) for the field (a) with $\sigma=1$ and $\varepsilon=10^{-4}$. Pronounced resonances occur in the vicinity of the electric nodal points owing to the $\varepsilon$-small magnetic-dipole response of the particle, i.e., its strong dual asymmetry.}
\label{fig:flowers}
\end{figure}

\textit{Conclusion.---}
We have examined a generic dipole interaction of a monochromatic electromagnetic field with bi-isotropic nanoparticles. Such interaction is described by the terms related to the breaking of dual, P- and T-symmetries, including chirality and the nonreciprocal magnetoelectric effect (``false chirality''). Calculating the absorption rates, radiation forces, and radiation torques, we have introduced novel local characteristics of the field, which quantify the energy, momentum, and angular-momentum transfer in these specific interactions. In particular, using the PT-symmetric `magnetoelectric energy density', we have described a complex field, which offers ultra-sensitive probing of the nonreciprocal magnetoelectric effect in nanoparticles. This can lead to the discovery of the magnetoelectric effect at the molecular level, as was discussed by Curie and Debye almost a century ago \cite{14,Curie}. We also expect that the magnetoelectric energy can serve as an important characteristic of various PT-symmetric effects in electromagnetism~\cite{16,20}.

\begin{acknowledgements}
This work was partially supported by the RIKEN iTHES Project, MURI Center for Dynamic Magneto-Optics, JSPS-RFBR contract no. 12-02-92100, Grant-in-Aid for Scientific Research (S), MEXT Kakenhi on Quantum Cybernetics, the JSPS via its FIRST program, and the Australian National University.
\end{acknowledgements}


\begin{thebibliography}{99}
\bibitem{remark} Rigorously speaking, monochromatic fields are not square-integrable functions, but this does not affect our local analysis.
We also note that the quantity (1) is not a true state vector (the electric and magnetic fields have different symmetry properties), but formally using the representation (1) facilitates calculations and illuminates the picture of electromagnetic interactions.
\bibitem{remark II} Because of the strong dual asymmetry, the fields are often described by the {\it electric} characteristics only (e.g., ${\bf{P}} \to 2{{\bf{P}}^{\rm (e)}}$, etc.), but these are not fundamental characteristics of free Maxwell fields \cite{Berry,9,10}.
\bibitem{29} 	Note that the resonant magnetoelectric response, Eq.~(25) and Fig.~1, cannot be mixed up with the resonant chiral response in \cite{6,Cho}, because the latter vanishes for equal amplitudes of the counter-propagating waves, and also has a symmetric rather than anti-symmetric profile in the vicinity of the nodal points (Fig.~1b).
\end{thebibliography}
\end{document}